\documentclass[twocolumn,showpacs,preprintnumbers,prl]{revtex4}
\usepackage{graphicx,epsfig}

\begin{document}

\preprint{PITHA 03/13; IPPP/03/82; FERMILAB-Pub-03/407-T; hep-ph/0312331}

\title{
\boldmath
Effective theory approach to unstable particle production
\unboldmath}
\author{M.~Beneke${}^1$, A.P.~Chapovsky${}^{1}$,
A. Signer${}^2$ and
G. Zanderighi${}^{\,3}$}

\affiliation{
$^1\!\!\!$ Institut f\"ur Theoretische Physik E, RWTH Aachen,
D-52056 Aachen, Germany\\
$^2\!\!\!$ IPPP, Department of Physics, University of Durham,
Durham DH1 3LE, England \\
$^3\!\!\!$ Fermi National Accelerator Laboratory, Batavia, IL 60510
}

\date{December 22, 2003}

\begin{abstract}
\noindent
Using the hierarchy of scales between the mass, $M$, and the width,
$\Gamma$, of a heavy, unstable particle we construct an effective
theory that allows calculations for resonant processes to be
systematically expanded in powers of the coupling $\alpha$
and $\Gamma/M$. We illustrate the method by computing the
next-to-leading order line shape of a scalar resonance in an abelian
gauge-Yukawa model.
\end{abstract}

\maketitle

Higher-order calculations for processes involving massive, unstable
particles close to resonance suffer from the breakdown of ordinary
perturbation theory, since the intermediate propagator becomes
singular. This singularity is avoided if the finite width, $\Gamma$,
of the unstable particle is taken into account in the construction of
the propagator via resummation of self-energy insertions. There are a
number of approaches along this line to avoid the
problem~\cite{unstable}. However, so far there is no method that
allows to systematically improve the accuracy of calculations
order by order in perturbation theory. The purpose of this letter is
to present such a method.

We are concerned with processes involving an unstable particle close
to resonance. The main idea is to exploit the hierarchy of scales
$\Gamma\ll M$, where $M$ is the pole mass, in order to systematically
organize the calculations in a series in the coupling, $\alpha$, and
$\Gamma/M$. While the expansion in $\alpha$ is standard, we construct
an effective theory to perform the expansion in $\Gamma/M$. A first
step in this direction has been presented in \cite{effective}. The
main idea of our approach is similar to non-relativistic QCD, where an
expansion in $\alpha$ and the velocity of the heavy quarks is made. We
will identify all relevant modes and use them to write the operators
of the Lagrangian of the effective theory. This Lagrangian is then
matched to the underlying theory, using the method of regions
\cite{Beneke:1998zp}. In this letter we will outline the basic idea and we
refer to \cite{BCSZ} for more details.

We consider a toy model that involves a massive scalar field, $\phi$,
and two fermion fields. The scalar as well as one of the fermion
fields, $\psi$, (the ``electron'') are charged under an abelian gauge
symmetry, whereas the other fermion, $\chi$, (the ``neutrino'') is
neutral. The model allows for the scalar to decay into an
electron-neutrino pair through a Yukawa interaction. The Lagrangian is
\begin{eqnarray}
\label{model}
{\cal L} &=& (D_\mu\phi)^\dagger D^\mu\phi - \hat M^2 \phi^\dagger\phi +
 \bar\psi i \!\not\!\!D\psi + \bar\chi i\!\!\not\!\partial\chi
\nonumber \\
&& - \, \frac{1}{4} F^{\mu\nu}F_{\mu\nu}-\frac{1}{2\xi} \,
(\partial_\mu A^\mu)^2
 \nonumber\\
 && + \, y\phi\bar\psi\chi + y^* \phi^\dagger \bar\chi\psi
-\frac{\lambda}{4}(\phi^\dagger \phi)^2+ {\cal L}_{\rm ct}\, ,
\end{eqnarray}
where $\hat{M}$ and ${\cal L}_{\rm ct}$ denote the renormalized mass and the
counterterm Lagrangian and $D=\partial-i g A$. We define $\alpha_g\equiv
g^2/(4\pi)$, $\alpha_y\equiv (y y^*)/(4 \pi)$ (at the scale $\mu$) and assume
$\alpha_g \sim \alpha_y \sim \alpha$, and $\alpha_\lambda\equiv\lambda/(4\pi)
\sim \alpha^2/(4\pi)$.

We would like to obtain the totally inclusive cross section for the
process
\begin{equation}
\bar\nu(q) + e^-(p)\to X
\label{process}
\end{equation}
as a function of $s\equiv (p+q)^2$ by calculating the forward
scattering amplitude ${\cal T}(s)$ and taking its imaginary part. (The
total cross section of process (\ref{process}) has an initial state
collinear singularity which has to be absorbed into the electron
distribution function. In what follows it is understood that this
singularity is subtracted minimally.)  In particular, we are
interested in the region $s\approx M^2$, or more precisely $s-M^2 \sim
M\Gamma\sim\alpha M^2 \ll M^2$. In this kinematic region the cross
section is enhanced due to the propagator of the scalar. Furthermore,
at each order in $\alpha$ we get additional contributions proportional
to $\alpha \hat{M}^2/(s-\hat{M}^2) \sim 1$ due to self-energy
insertions.

We now turn to the main part of this letter and discuss how to
construct the effective theory. Our approach is based
on the hierarchy of scales $\Gamma\ll M$. Thus, we systematically
expand the cross section in powers of $\alpha$ and
\begin{equation}
\label{eq:deltadef}
\delta\equiv \frac{s-\hat M^2}{\hat M^2} \sim \frac{\Gamma}{M} .
\end{equation}
In a theory that formulates this expansion correctly, other issues
like resummation of self-energy insertions and gauge invariance are
taken care of automatically. In a first step we integrate out hard
momenta $k\sim M$. The effective theory will then not contain any
longer dynamical hard modes since their effect is included in the
coefficients of the operators. The hard effects are associated with
what is usually called factorizable corrections, whereas the effects
of the dynamical modes correspond to the non-factorizable corrections
\cite{effective}. On the level of Feynman diagrams, this amounts to
using the method of regions to separate loop integrals into various
contributions~\cite{Beneke:1998zp}. The hard part is obtained by
expanding the integrand in $\delta$. The difference between the full
integral and its hard part has to be reproduced by modes corresponding
to momentum configurations that are near mass-shell. The main task is
to identify these modes, and to write the operators of the effective
Lagrangian in terms of the corresponding field operators and then to
compute the coefficients of the operators by matching (up to a certain
order in $\alpha$ and $\delta$).

Our goal is to carry out this programme for our model to an order in
$\alpha$ and $\delta$ that is sufficient to compute ${\cal T}^{(0)}
+{\cal T}^{(1)}$, the forward scattering amplitude at next-to-leading
order (NLO), where ${\cal T}^{(0)}$ sums up all terms that scale as
$(\alpha/\delta)^n\sim 1$ and ${\cal T}^{(1)}$ contains all terms that
are suppressed by an additional power of $\alpha$ or $\delta$.

The basic process under consideration is the following: we start with
highly energetic fermions, produce a near mass-shell scalar which then
decays again into highly energetic fermions. Accordingly we split the
effective Lagrangian into three parts.  Roughly speaking, the first,
${\cal L}_{\rm HSET}$, describes the heavy scalar field near
mass-shell and its interaction with the gauge field. The second part,
${\cal L}_{\rm SCET}$, describes energetic (charged) fermions and
their interactions with the gauge field. Finally, the third part,
${\cal L}_{\rm int}$, describes the external fermions and how they
interact to produce the final state. We will discuss these three parts
in turn.

The construction of ${\cal L}_{\rm HSET}$ follows closely the
construction of the effective Lagrangian for heavy quark effective
theory (HQET) \cite{Eichten:1990zv}. We write the momentum of the
scalar particle near resonance as $P=\hat{M}v + k$, where the velocity
vector $v$ satisfies $v^2=1$ and the residual momentum $k$ scales as
$M\delta$. We will call such a scalar field a ``soft'' field (in
\cite{effective} the term ``resonant'' has been used).  Thus, for a
soft scalar field we have $P^2-\hat{M}^2 \sim M\delta$ and this
remains true if the scalar particle interacts with a soft gauge boson
with momentum $M\delta$. In analogy to HQET we remove the rapid
spatial variation $e^{-i\hat{M} v\cdot x}$ from the $\phi$-field and
define
\begin{equation}
\label{phivdef}
\phi_v(x) \equiv e^{i\hat{M} v\cdot x}\, {\cal P}_+ \phi(x)\, ,
\end{equation}
where ${\cal P}_+$ projects onto the positive frequency part to ensure
that $\phi_v$ is a pure destruction field.  We now write the effective
Lagrangian in terms of $\phi_v$ and construct the bilinear terms so as
to reproduce the two-point function close to resonance. Denoting the
complex pole of the propagator by $\bar{s}$ and the residue at the
pole by $R_\phi$ the propagator can be written as
\begin{equation}
\frac{i\, R_\phi}{P^2 - \bar s} =
\frac{i\, R_\phi}{2 \hat{M} v k + k^2 -(\bar s -\hat{M}^2)}.
\label{eq:propagator}
\end{equation}
We define the matching coefficient
\begin{equation}
\Delta \equiv \frac{\bar s -\hat{M}^2}{\hat{M}}
\label{eq:Deltadef}
\end{equation}
and $a^\mu_\top\equiv a^\mu-(va)\, v^\mu$ for any vector. There are two
solutions to $P^2=\bar{s}$, one of which is irrelevant since it scales as
$vk\sim\hat{M}$. For the other we find
\begin{eqnarray}
vk &=& -\hat{M}+\sqrt{\hat{M}^2+\hat{M}\Delta-k_\top^2} \nonumber \\
&=& \frac{\Delta}{2}-\frac{\Delta^2+4 k_\top^2}{8\hat{M}}
 + {\cal O}(\delta^3) ,
\label{propexp}
\end{eqnarray}
where we expanded in $\delta$ in the second line.
Therefore, the bilinear terms are given by
\begin{eqnarray}
\label{eq:Leff_phiphi2}
  {\cal L}_{\phi\phi} &=&
 2 \hat{M}  \phi_v^\dagger\,
        \left(iv\cdot D_s  - \frac{\Delta}{2} \right) \phi_v
 \nonumber \\
&+& 2 \hat{M}  \phi_v^\dagger\,
        \left( \frac{(i D_{s\top})^2}{2\hat{M}} +
               \frac{\Delta^2}{8 \hat{M}} \right) \phi_v
 + \ldots,
\end{eqnarray}
where $D_s\equiv \partial -ig A_s$ denotes the soft covariant
derivative. In obtaining ${\cal L}_{\phi\phi}$ we exploited the fact
that the gauge invariance of the full Lagrangian is not broken by the
separation into hard and soft parts. Therefore, the effective
Lagrangian must be gauge invariant as well and we can obtain the
interaction of the scalar with the soft photon simply by replacing
$\partial\to D_s$. The gauge invariance of $\Delta$ follows from the
gauge invariance of $\bar{s}$ and $\hat{M}$. Furthermore, $\Delta$ is
given entirely by hard contributions, which justifies its
interpretation as matching coefficient. Using (\ref{eq:Deltadef}) we
can express it in terms of the hard part of the self-energy
$\Pi_h(s)$. Writing $\Pi_h(s) = \hat{M}^2\, \sum_{k,l} \delta^l \,
\Pi^{(k,l)}$, where it is understood that $\Pi^{(k,l)}\sim\alpha^k$,
we obtain
\begin{eqnarray}
\lefteqn{\Delta \equiv \sum_i  \Delta^{(i)} = }&&
\label{eq:Deltaexp} \\
&&  \hat{M}\, \Pi^{(1,0)} +
 \hat{M}  \left(\Pi^{(2,0)}+\Pi^{(1,1)}\Pi^{(1,0)} \right) + \ldots
\nonumber
\end{eqnarray}
Explicit results for $\Delta^{(1)}$ and $\Delta^{(2)}$ in the
$\overline{\rm MS}$ and pole renormalization scheme can be found in
\cite{BCSZ}. Here we only note that in the pole scheme $\bar s \equiv
M^2-i M \Gamma$, so $\Delta = -i\Gamma$ when $\hat{M} = M$. Inserting
the expansion (\ref{eq:Deltaexp}) into (\ref{eq:Leff_phiphi2}) and
supplementing ${\cal L}_{\phi\phi}$ with the kinetic terms for soft
photons and fermions we obtain
\begin{eqnarray}
{\cal L}_{\rm HSET} &=&  2 \hat{M}  \phi_v^\dagger\,
        \left( i v \cdot D_s - \frac{\Delta^{(1)}}{2} \right) \phi_v
\nonumber \\
&+& 2 \hat{M}  \phi_v^\dagger\,
        \left( \frac{(i D_{s,\top})^2}{2\hat{M}} +
               \frac{[\Delta^{(1)}]^2}{8\hat{M}} -
               \frac{\Delta^{(2)}}{2} \right) \phi_v
\nonumber  \\
&-&\frac{1}{4}\,F_{s\mu\nu} F_s^{\mu\nu}
+\bar\psi_s i\!\not\!\!D_s \psi_s
+\bar\chi_s i\!\not\!\partial \chi_s .
\label{heavyNLO}
\end{eqnarray}
Each term in ${\cal L}_{\rm HSET}$ can be assigned a scaling power in
$\delta$. In momentum space the propagator of the $\phi_v$ field
scales as $1/\delta$. Hence, because $\int d^4 k$ counts as
$\delta^4$, the soft scalar field $\phi_v(x)$ scales as
$\delta^{3/2}$. Since $\Delta^{(1)}\sim D_s \sim M\delta$, both terms
in the first line of (\ref{eq:Leff_phiphi2}) scale as $\delta^4$ and
are leading terms. The terms in the second line are suppressed by one
power in $\delta$ or $\alpha$. Finally, since $A^\mu_s$ scales as
$\delta$ and the soft fermion fields scale as $\delta^{3/2}$ (see
\cite{BCDF}) the terms in the last line of (\ref{eq:Leff_phiphi2})
scale as $\delta^4$. In (\ref{eq:Leff_phiphi2}) we have left out terms
further suppressed in $\delta$ or $\alpha$. As we will see, they are
not needed for the calculation of the line shape at NLO. However, we
stress that the expansion can be performed to whatever accuracy is
needed.

We note that computing the scalar propagator to all orders in $\delta$
using ${\cal L}_{\rm HSET}$ does not reproduce
(\ref{eq:propagator}). Instead near resonance we obtain $i \varpi^{-1}
R_{{\rm eff}\phi}/(P^2-\bar{s})$, where $\varpi^{-1}\equiv
(\hat{M}^2+\hat{M}\Delta -k_\top^2)^{1/2}/\hat{M} = 1+{\cal
O}(\delta,\alpha)$. The difference in the normalization is taken into
account in matching calculations by an additional wave-function
normalization factor $\varpi^{-1/2}$ for each external $\phi_v$-line
in the effective theory.

Next, we turn to the construction of the effective Lagrangian, ${\cal
L}_{\rm SCET}$, associated with the energetic fermions. We need a
``collinear'' mode to describe a fermion with large momentum in the
say $\vec{n}_-$ direction. Such modes have been discussed previously
within the context of soft-collinear effective theory (SCET)
\cite{Bauer:2000ew}. The Lagrangian has been worked out to order
$\delta$ in \cite{BCDF} and we can take the parts relevant to us from
there. (What we call ``soft'' here what is usually called
``ultrasoft'' in the context of SCET and in the power counting our
$\delta$ corresponds to $\lambda^2$ in
\cite{BCDF}.) For each direction defined by an energetic particle we
introduce two reference light-like vectors, $n_\pm$, with
$n_+^2=n_-^2=0$ and $n_+ n_-=2$ and we write the corresponding
momentum as
\begin{equation}
  p^\mu = (n_+p) \, \frac{n_-^\mu}{2} + p_\perp^\mu + (n_-p)
  \frac{n_+^\mu}{2},
\label{collmom}
\end{equation}
where $n_+ p \sim M$, $n_- p \sim M\delta$ and
$p_\perp \sim M\delta^{1/2}$. Given a certain direction $n_-$ we
introduce the collinear field $\psi_c$ which satisfies $\not\!\!n_-
\psi_c = 0$. The terms relevant for the calculation of ${\cal
T}^{(0)}+{\cal T}^{(1)}$ are then given by
\begin{equation}
\label{scetterm}
{\cal L}_{\rm SCET} =
\bar{\psi}_c \left(i n_- D +  i \!\not\!\!D_{\perp c}
\frac{1}{i n_+ D_{c}+i\epsilon}\, i\!\not\!\!D_{\perp c} \right)
\frac{\not\!n_+}{2} \, \psi_c\, .
\end{equation}
Since we are concerned with the forward scattering amplitude, the only
directions defined by energetic particles are given by the incoming
electron and (anti)neutrino. Thus, we have two sets of collinear
modes, one for the incoming electron, $\psi_{c1}$, and one for the
incoming (anti)neutrino, $\chi_{c2}$. Of course, in the case of the
neutrino, the covariant derivatives in (\ref{scetterm}) have to be
replaced by ordinary derivatives.  All terms in (\ref{scetterm}) scale
as $\delta^2$. Terms of order $\delta^{5/2}$ and $\delta^3$ exist, but
they are not needed for our application, since they would result in
contributions suppressed by an additional power of $\alpha$ and,
therefore, contribute only at NNLO. Again there is no difficulty in
going to higher orders in the expansion if needed.

The last part to consider is ${\cal L}_{\rm int}$. It has to include
operators that allow the production and decay of the unstable
particle. Without introducing additional modes it is not possible to
include such vertices as ordinary interaction terms in the effective
Lagrangian \cite{BCSZ}. The reason is that the momenta associated with
generic collinear fields $\psi_{c1}$ and $\bar{\chi}_{c2}$ do not add
up to a momentum of the form $P=M v+k$. Either we have to implement
this kinematic constraint on our external states by hand \cite{BCSZ}
or we have to introduce a new ``external-collinear'' mode. Adopting
the second option, we define an external-collinear mode with large
momentum in the $\vec{n}_-$ direction by assigning it a momentum
$\hat{M} n_-/2+k$, where $k\sim\delta$. This mode has the same
virtuality $\hat{M} \delta^{1/2}$ as a generic collinear mode but the
momentum is not given by (\ref{collmom}), because it has a fixed large
component such that the two incoming fermions produce a scalar near
mass shell. For such a mode it is useful to extract the fixed large
momentum and to define
\begin{equation}
\psi_{n_-}(x) \equiv e^{i\hat{M}/2\,(n_- x)}\, {\cal P}_+\,
\psi_{c1}(x) ,
\end{equation}
and similarly for $\chi_{n_+}$. For the purpose of computing ${\cal
T}^{(0)}+{\cal T}^{(1)}$ it is sufficient to take the first term of
${\cal L}_{\rm SCET}$, (\ref{scetterm}), with a soft photon only to
describe the interaction of the external-collinear fermions with the
photons
\begin{equation}
{\cal L}_{\pm} =
\bar{\psi}_{n_-} \!i n_- D_s
\frac{\!\not\!n_+}{2} \, \psi_{n_-} +
\bar{\chi}_{n_+} \! i n_+ \partial\,
\frac{\!\not\!n_-}{2} \, \chi_{n_+} .
\label{Lpm}
\end{equation}
With the external-collinear modes we can implement the production and
decay vertices as interaction terms in ${\cal L}_{\rm int}$. It is
also convenient to integrate out generic collinear fields and keep
only the external-collinear modes in the effective theory. Because
adding soft fields results in a further suppression in $\delta$ we
then find that we can restrict ourselves to
\begin{eqnarray}
{\cal L}_{\rm int} &=& C\, y\, \phi_v \bar{\psi}_{n_-}\chi_{n_+}
+ C\, y^* \phi_v^\dagger \bar{\chi}_{n_+}\psi_{n_-}
\nonumber \\
&+& F\, \frac{y y^*}{\hat{M}^2}
  \left(\bar{\psi}_{n_-} \chi_{n_+}\right)  \!
  \left(\bar{\chi}_{n_+} \psi_{n_-}\right) ,
\label{eq:Lint}
\end{eqnarray}
where $C=1+{\cal O}(\alpha)$ and $F$ are the matching
coefficients. The external fields scale as $\delta^{3/2}$. Thus, an
insertion of a $\phi\psi\chi$ operator results in $\int d^4x\, \phi_v
\bar{\psi}_{n_-}\chi_{n_+} \sim \delta^{1/2}$. The forward scattering
amplitude can be obtained by two insertions of this operator. Taking
into account the scaling of the external state $\langle \bar{\nu} e^-|
\sim \delta^{-1}$ we see that ${\cal T}^{(0)} \sim \alpha/\delta$.
The four-fermion operator is suppressed in $\delta$ and results in a
contribution of order $\alpha$ to ${\cal T}$. Thus, to compute ${\cal
T}^{(1)}$ we need $C^{(1)}$, the ${\cal O}(\alpha)$ contribution to
the matching coefficient $C$, while $F$ is only needed at tree level.

The coefficient $C^{(1)}$ is obtained by matching the on-shell
three-point function of a scalar field, an electron and a neutrino at
order $y \alpha$ and at leading order in $\delta$. In particular, this
involves the computation of (the hard part) of the vertex diagram, and
the additional wave-function normalization factor $\varpi^{-1/2}$
mentioned above has to be taken into account. For the precise matching
equation as well as the explicit result for $C^{(1)}$ we refer to
\cite{BCSZ}. Here it suffices to say that these are standard loop
calculations. To obtain $F^{(0)}$ (the LO contribution to $F$) we have
to match the four-point function at tree level, but include subleading
terms in $\delta$. The explicit result is $F^{(0)} = 1/4$.

We have now completed the construction of the effective Lagrangian
${\cal L}_{\rm eff} = {\cal L}_{\rm HSET} + {\cal L}_{\pm} + {\cal
L}_{\rm int}$ to an accuracy  sufficient to compute ${\cal T}$ at
NLO. At leading order there is only one diagram, involving two
three-point vertices and one resonant scalar propagator. We get
\begin{equation}
\label{LOT0}
i {\cal T}^{(0)} =
\frac{- i\, y y^*}{2\hat M {\cal D}} \,
[\bar{u}(p)v(q)]\,[\bar{v}(q)u(p)] ,
\end{equation}
where we defined ${\cal D}\equiv \sqrt{s}-\hat M-\Delta^{(1)}/2$. In
the effective theory there are three classes of diagrams that
contribute to ${\cal T}^{(1)}$. Firstly, there are hard corrections
consisting of a propagator insertion $[\Delta^{(1)}]^2/4 - \hat{M}
\Delta^{(2)}$ as well as a vertex insertion $C^{(1)}$. Secondly, there
is a four-point vertex diagram due to the
$(\bar{\psi}\chi)(\bar{\chi}\psi)$ operator in ${\cal L}_{\rm
int}$. The third class are soft-photon loop diagrams, corresponding to
the non-factorizable corrections. Adding up all these contributions
and using the explicit result for $C^{(1)}$ (in the $\overline{\rm
MS}$ scheme) \cite{BCSZ} we obtain
\begin{eqnarray}
\lefteqn{i\,{\cal T}^{(1)} = i\,{\cal T}^{(0)} \times} &&
\label{eq:T1finite} \\
&&  \Bigg[\
a_g \left( 3 \ln\frac{-2\hat{M}{\cal D}}{\nu^2}
   + 4\, \ln\frac{-2\hat{M} {\cal D}}{\hat{M}^2}
     \ln\frac{-2\hat{M}{\cal D}}{\nu^2}
    \right.
\nonumber \\
&&  \left. \qquad -\ 7 \ln\frac{-2\hat{M} {\cal D}}{\hat{M}^2}
           - \frac{3}{2} \ln\frac{\hat{M}^2}{\mu^2}
           -\frac{7}{2} + \frac{2\pi^2}{3} \right)
\nonumber \\
&& +\ a_y \left( 2 \ln\frac{\hat{M}^2}{\mu^2}
                 -\frac{1}{2} -i\pi \right)
   - \frac{[\Delta^{(1)}]^2}{8{\cal D}\hat{M}}
   + \frac{\Delta^{(2)}}{2{\cal D}} - \frac{{\cal D}}{2\hat{M}} \ \Bigg],
\nonumber
\end{eqnarray}
where $a_i\equiv \alpha_i/(4\pi)$. The initial state collinear
singularities have been subtracted minimally and we denote the
corresponding factorization scale by $\nu$ to distinguish it from the
renormalization scale $\mu$.

\begin{figure}
\includegraphics[width=0.45\textwidth]{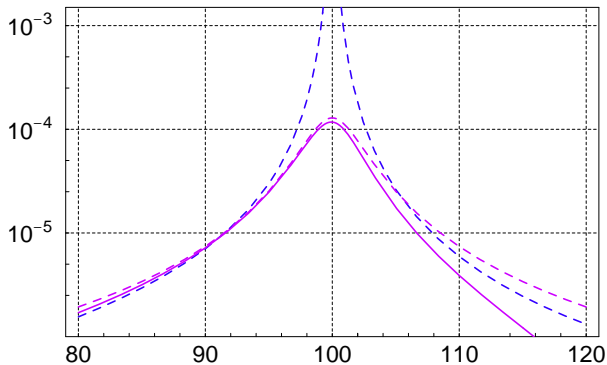}
\caption{\label{fig:ls} The line shape (in GeV$^{-2})$ in the
effective theory at LO (light grey/magenta dashed) and NLO (light
grey/magenta) and the LO cross section off resonance in the full
theory (dark grey/blue dashed) as a function of the center-of-mass
energy (in GeV).}
\end{figure}

We can now perform the polarization average and take the imaginary
part of $({\cal T}^{(0)}+{\cal T}^{(1)})/s$. This result describes the
line shape near resonance with a relative error of $\alpha^2$. Moving
away from the resonance, the relative error becomes of order unity,
since $\delta$ is not small any longer. To obtain a good description
for all values of $\sqrt{s}$, the result of the effective theory has
to be matched to the off-resonance result of the full theory.

In Figure~\ref{fig:ls} we show the leading order line shape in the
effective theory and the tree-level (order $\alpha^2$) cross section
off resonance in the full theory. The two results agree in an
intermediate region where both calculations are valid. This allows to
obtain a consistent LO result for all values of $\sqrt{s}$. We also
show the NLO line shape. For the numerical results we have chosen to
use the $\overline{\rm MS}$ scheme with $\alpha_y=\alpha_g=0.1$ and
$\alpha_\lambda=(0.1)^2/(4\pi)$. The pole mass is assumed to be
$M=100$~GeV which results in the $\overline{\rm MS}$ value
$\hat{M}=98.8$~GeV for the LO result and $\hat{M}=99.1$~GeV for the
NLO result.  Furthermore, we have chosen a variable factorization
scale such that there are no large logarithms involving $\nu$. We
remark that in order to obtain an improved NLO result for the whole
region of $\sqrt{s}$, the NLO line shape would have to be matched to
the NLO off-resonance cross section in the full theory.

The example considered here is based on a rather simple toy
model. Nevertheless, it allows to address the conceptual issues
related to unstable particles. The main result is that, using an
effective theory approach, calculations can be performed in a
systematic way in expanding in the small quantities $\alpha$ and
$\Gamma/M$. Applying our method to the Standard Model might require
more tedious calculations, but the main result remains
valid. In particular, as discussed in \cite{BCSZ}, NNLO line-shape
calculations now appear feasible.

\begin{acknowledgments}
The work of M.B. and A.P.C. is supported in part by the DFG
Sonderforschungsbereich/Transregio 9 ``Computer-gest\"utzte
Theoretische Teilchenphysik''
\end{acknowledgments}

\end{document}